\documentclass[12pt,preprint]{aastex62}
\usepackage{booktabs, float, amsmath, rotating}
\def\lea{\mathrel{<\kern-1.0em\lower0.9ex\hbox{$\sim$}}}
\def\gea{\mathrel{>\kern-1.0em\lower0.9ex\hbox{$\sim$}}}

\graphicspath{{./}{./Figures/}{figures/}}
\received{}
\revised{}
\accepted{}
\submitjournal{ApJ}

\shorttitle{Mass-Radius Relation and Mass Functions of Molecular Clumps in the LMC}
\shortauthors{Mok et al.}

\def\aap{A\&A}

\def\apjl{ApJ}

\def\apjs{ApJS}

\begin{document}

\title{Feedback in Forming Star Clusters: The Mass-Radius Relation and Mass Function of Molecular Clumps in the Large Magellanic Cloud \vspace{-2cm}}
\correspondingauthor{Rupali Chandar}
\email{Rupali.Chandar@utoledo.edu}
\author[0000-0001-7413-7534]{Angus Mok}
\affil{Department of Physics \& Astronomy, The University of Toledo, Toledo, OH 43606, USA}
\author{Rupali Chandar}
\affil{Department of Physics \& Astronomy, The University of Toledo, Toledo, OH 43606, USA}
\author{S. Michael Fall}
\affil{Space Telescope Science Institute, Baltimore, MD 21218, USA}

\doublespace
\begin{abstract}
We derive the mass-radius relation and mass function of molecular clumps in the Large Magellanic Cloud (LMC) and interpret them in terms of the simple feedback model proposed by Fall, Krumholz, and Matzner (FKM). Our work utilizes the dendrogram-based catalog of clumps compiled by Wong et al. from $^{12}$CO and $^{13}$CO maps of six giant molecular clouds in the LMC observed with the Atacama Large Millimeter Array (ALMA). The Magellanic Clouds are the only external galaxies for which this type of analysis is possible at the necessary spatial resolution ($\sim1$ pc). We find that the mass-radius relation and mass function of LMC clumps have power-law forms, $R \propto M^{\alpha}$ and $dN/dM \propto M^{\beta}$, with indices $\alpha = 0.36 \pm 0.03$ and $\beta= -1.8 \pm 0.1 $ over the mass ranges $10^2 M_\odot \la M \la 10^5 M_\odot$ and $10^2 M_\odot \la M \la 10^4 M_\odot$, respectively. With these values of $\alpha$ and $\beta$ for the clumps (i.e., protoclusters), the predicted index for the mass function of young LMC clusters from the FKM model is $\beta \approx 1.7$, in good agreement with the observed index. The situation portrayed here for clumps and clusters in the LMC replicates that in the Milky Way.
\end{abstract}
\section{Introduction} \label{sec:intro}
\par
Star clusters form in the dense subunits of giant molecular clouds (GMCs) known as clumps \citep{Lada03, McKee07, Krumholz19}. Thus, the properties of clusters must reflect the properties of clumps (as ``initial’’ conditions) modified by the actions of star formation and stellar feedback within them. In this context, two of the most relevant statistical properties of cluster and clump populations are the mass function and mass--radius relation, usually represented by power laws: $\psi(M) \equiv dN/dM \propto M^{\beta}$ and $R \propto M^{\alpha}$. A comparison of the indices $\beta$ and $\alpha$ for clusters and clumps should then tell us something about star formation and feedback. \citet*[][hereafter FKM]{Fall10} developed this idea into a simple analytical model and applied it to observations of clusters and clumps in the Milky Way. In a previous paper, we applied the model to observations of clusters and GMCs in six nearby galaxies (\citealt{Mok20}; see also \citealt{Hughes13b} for a similar study of M51). Here, we apply the FKM model to observations of clusters and clumps in the Large Magellanic Cloud (LMC) for the first time.
\par
The mass functions of clusters, clumps, and GMCs are related by the star formation efficiencies (SFEs) $\mathcal{E}_{\rm clump} \equiv M_{\rm cluster} / M_{\rm clump}$ and $\mathcal{E}_{\rm GMC} \equiv M_{\rm cluster} / M_{\rm GMC}$. Most determinations of these mass functions find power laws with indices $\beta \approx -2$ within a range of $\pm 0.3$ (for clusters, see \citealt{Zhang99, Bik03, Cook19, Mok19}; for clumps, see \citealt{Munoz07, Wong08, Schlingman11, Pekruhl13, Urquhart14, Moore15, Brunetti19}; for GMCs, see \citealt{Fukui10, Wong11, Rice16, Mok20}). This indicates that the typical (mean or median) SFEs in clumps and GMCs are roughly independent of mass, provided that the typical numbers of clusters per clump and GMC are also independent of mass. Estimates of the SFEs themselves are indirect and uncertain, with typical values $\mathcal{E}_{\rm clump} \sim 10$\%-30\% and $\mathcal{E}_{\rm GMC} \sim 1$\%-3\% \citep{Lada03, Grudic18}. The fact that these SFEs are small indicates that feedback is highly effective in removing gas from protoclusters before it is converted into stars. 
\par
The mass--radius relation determines the mass dependence of various properties of the objects, including their gravitational binding energy and force per unit mass ($E/M \propto G M / R$ and $F/M \propto G M / R^2$) and their mean surface and volume densities ($\Sigma \propto M / R^2$ and $\rho \propto M / R^3$). The first two of these are relevant in the present context because it is gravity that feedback must overcome to remove the gas from protoclusters. Estimates of the index of the mass--radius relation for clusters in multiple galaxies range from $\alpha \approx 0.1$ \citep{Larsen04} to $\alpha \approx 0.3$ \citep{Fall12, Krumholz19}. For clumps in the Milky Way, the only galaxy surveyed until now, the range is $0.3 \la \alpha \la 0.6$ (\citealt{Wong08}; FKM; \citealt{Roman-Duval10, Wu10, Urquhart18}). The mass--radius relations of GMCs in different galaxies have similar indices, $\alpha \approx 0.5$, but different normalizations \citep{Larson81, Wong11, Miville-Deschenes17, Sun18}. Indices near $\alpha = 0.5$ may raise suspicions about selection effects, but these are unlikely to cause major biases in the estimates above. This is because the clusters have surface brightnesses well above detection limits, the clumps are often selected for their star formation activity, not their gas surface density, and the GMCs observed in different galaxies with the same sensitivity have different mean surface densities.
\par
In this paper, we derive and interpret the mass function and mass--radius relation of clumps in the LMC. Our analysis is based on the recent catalog of CO-detected clumps derived by \citet{Wong19} from observations of six GMCs in the LMC with the Atacama Large Millimeter Array (ALMA). This dataset is especially valuable because the Magellanic Clouds are the only galaxies beyond the Milky Way for which it is possible to map the distribution of molecular gas on the clump scale ($R \sim 1$~pc) with current facilities. We summarize the \citet{Wong19} observations and catalog in Section~\ref{sec:obs} and derive the mass function and mass--radius relation in Section~\ref{sec:results}. We then interpret these results in terms of the FKM model in Section \ref{sec:interpretation} and present our main conclusions in Section \ref{sec:conclusion}. 
\section{Observations} \label{sec:obs}
\par
We use the published \citet{Wong19} clump catalog derived from new and archival ALMA maps of six widely separated GMCs in the LMC (called 30 Dor, PGC, N59C, A439, GMC104, and GMC1). The first two of these GMCs were observed in the $^{12}$CO(2-1) and $^{13}$CO(2-1) lines, while the last four were observed in the $^{12}$CO(1-0) and $^{13}$CO(1-0) lines. Wong et al. smoothed all these maps to a common angular resolution of $3.5\arcsec$ ($\sim0.8$ pc) before constructing their clump catalog. 
\par 
The \citet{Wong19} catalog is the largest, most uniform, and most representative (but incomplete) sample of clumps in the LMC. All these data have been processed and analysed in the same way. Two other GMCs in the LMC have also been mapped by ALMA in the $^{12}$CO and $^{13}$CO lines to construct samples of clumps \citep{Nayak16, Nayak18, Naslim18}. However, because these samples are based on different resolutions, sensitivities, and selection criteria, we do not attempt to combine them with the \citet{Wong19} sample in the work presented here.
\par 
The \citet{Wong19} clump catalog is based on the dendogram analysis developed by \citet{Rosolowsky08} and others. This procedure decomposes intensity maps into a nested hierarchy of structures known as leaves, branches, and trunks. The leaves are located at the local intensity peaks and contain no resolved substructure, the branches contain leaves and other branches, while the trunks are the largest contiguous structures. Since the leaves, branches, and trunks are all subunits within GMCs, we refer to them collectively as clumps. The dendogram analysis is designed to reflect the hierarchical structure of the interstellar medium (ISM), and its leaves are likely sites of cluster formation. It is a popular and proven method for analyzing intensity maps but not the only one (clumpfind being the main alternative). Figure~1 shows the $^{12}$CO intensity maps of the six GMCs, along with the dendogram-based leaves, branches, and trunks from the \citet{Wong19} catalog.
\begin{figure}[ht]
	\centering
	\includegraphics[height=6.5cm]{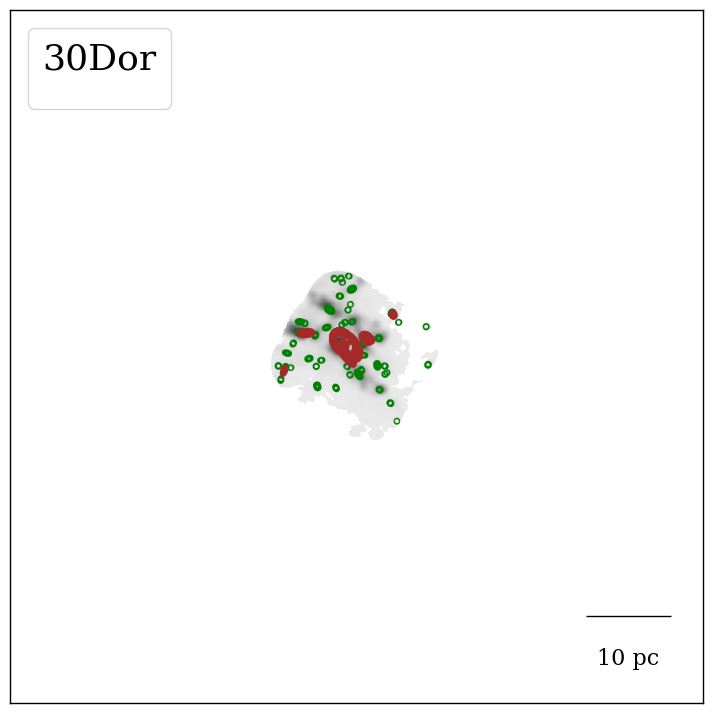}
	\includegraphics[height=6.5cm]{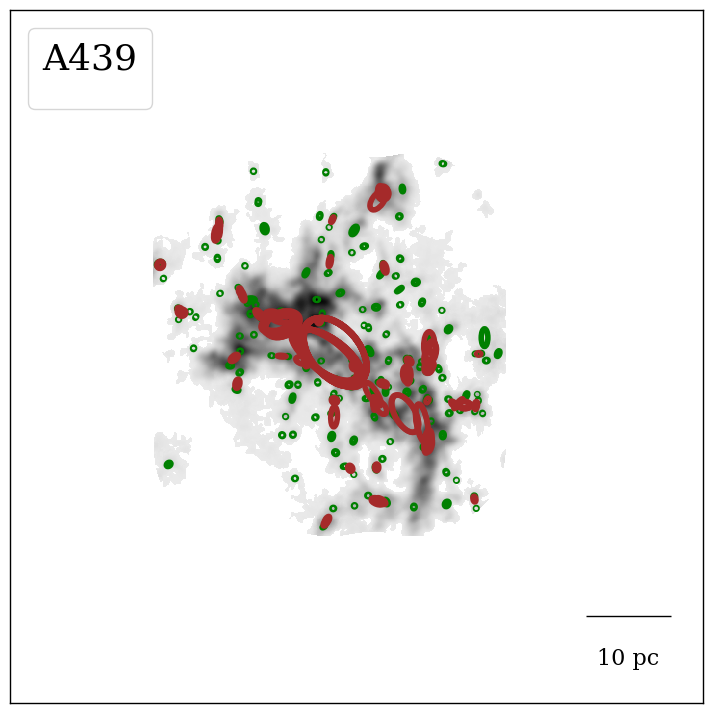}
	\includegraphics[height=6.5cm]{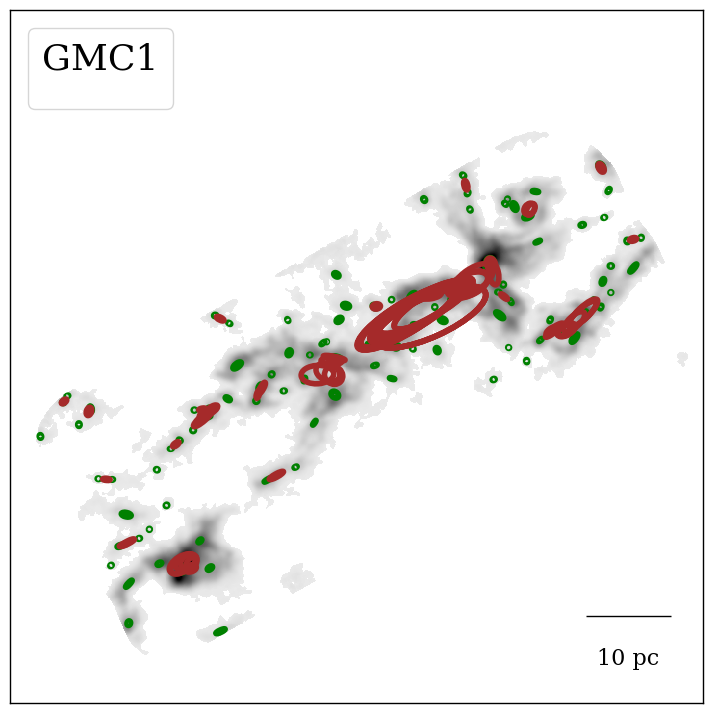}
	\includegraphics[height=6.5cm]{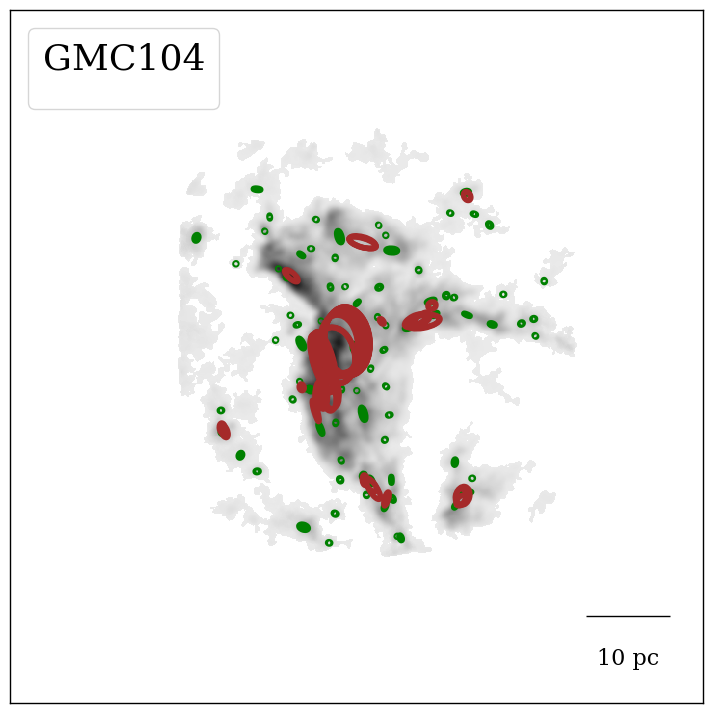}
	\includegraphics[height=6.5cm]{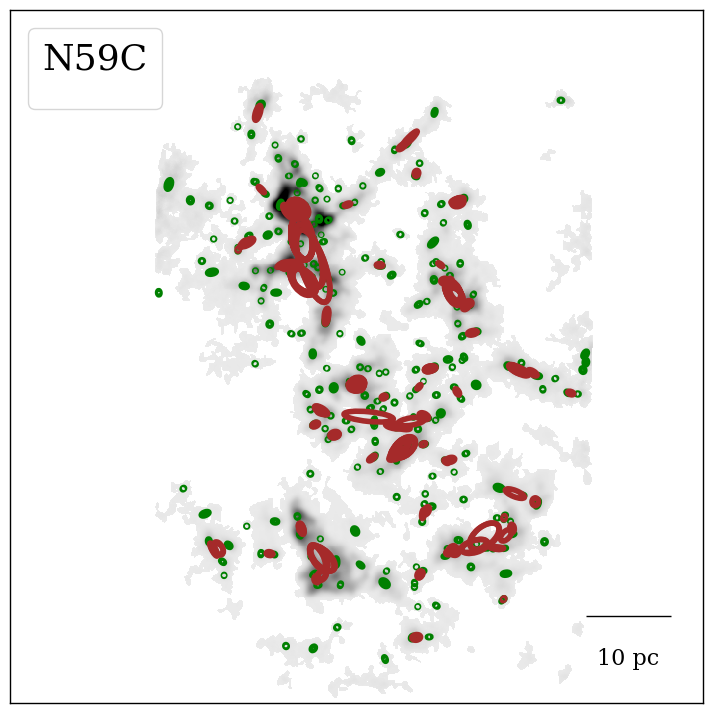}
	\includegraphics[height=6.5cm]{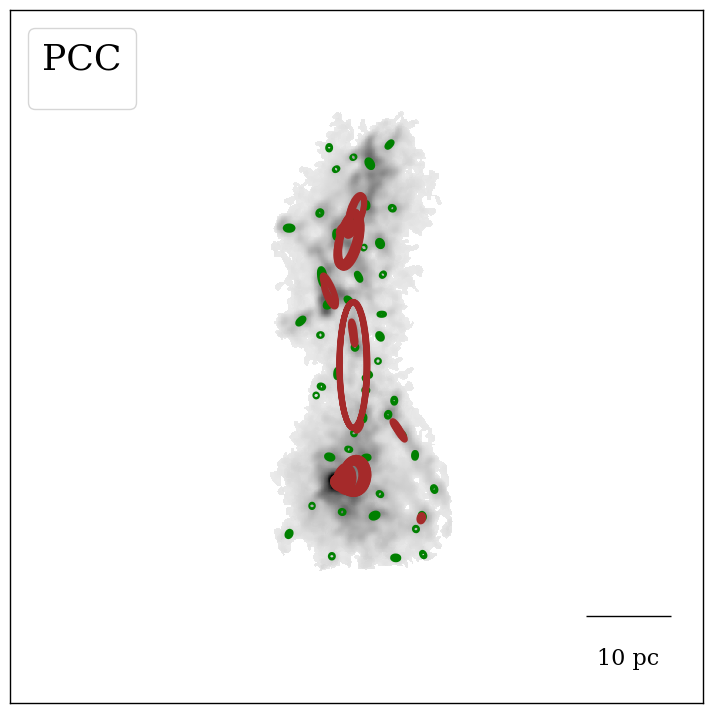}
	\caption{$^{12}$CO intensity maps (grayscale) of the six GMCs included in this study. The leaves (green) and branches plus trunks (brown) are represented by ellipses with dimensions and orientations derived by \citet{Wong19}. The panels are all aligned with North up and East left with the same angular scale in both directions, indicated by the inset bars. Note the hierarchical arrangement of clumps within these GMCs.}\label{fig-wong19}
\end{figure}
\par
The \citet{Wong19} catalog lists the masses $M$ and radii $R$ separately for clumps detected in the $^{12}$CO lines (which traces bulk molecular gas) and in the fainter, optically thin $^{13}$CO lines. The masses are estimated in three different ways: (1) $M_{\rm Lum}$ from the measured $^{12}$CO luminosity and an assumed $^{12}$CO-to-H$_2$ conversion factor\footnote{$M = \alpha_{\rm CO}\ L_{\rm CO}$, with $\alpha_{\rm CO}$ = $10.32\ M_\odot / ({\rm K\ km\ s^{-1}\ pc^{2}})$ for $^{12}$CO(1-0) and $\alpha_{\rm CO}$ = $12.9\ M_\odot / ({\rm K\ km\ s^{-1}\ pc^{2}}))$ for $^{12}$CO(2-1)}, (2) $M_{\rm LTE}$ from the measured $^{12}$CO and $^{13}$CO luminosities, the assumption of local thermodynamic equilibrium (LTE), and an assumed $^{13}$CO abundance ratio, and (3) $M_{\rm Vir}$ from the measured radius and velocity dispersion and the assumption of virial equilibrium. The radii are taken to be the geometric means of the major and minor axes of the CO contours. Previous studies have found that the exact method used to measure the sizes of clumps is unimportant, with similar size-linewidth relations found when different segmentation algorithms were applied to the same observations \citep{Colombo15}.
\par
\begin{figure}[ht]
	\centering
	\includegraphics[width=9.45cm]{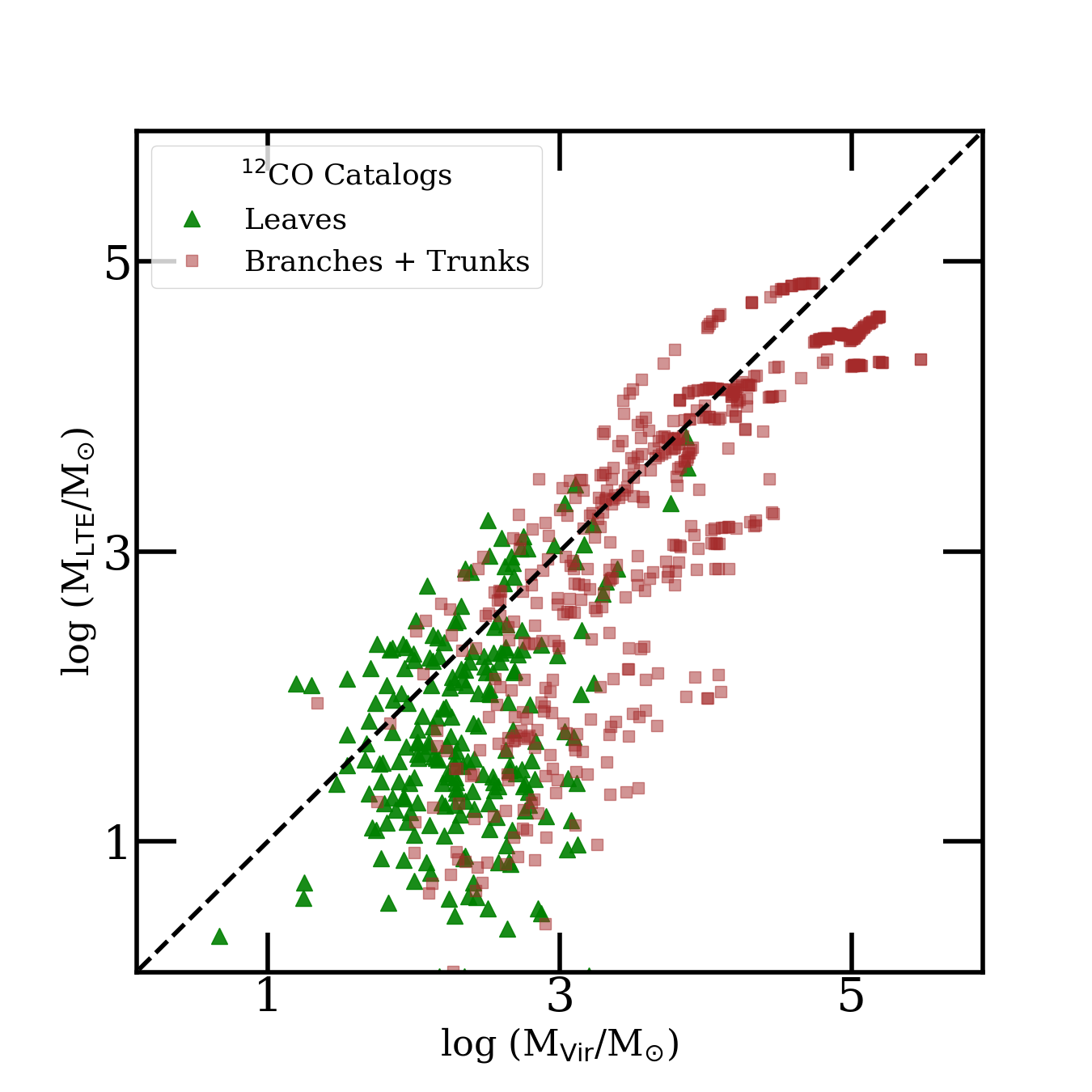} \hspace{-1.15cm}
	\includegraphics[width=9.45cm]{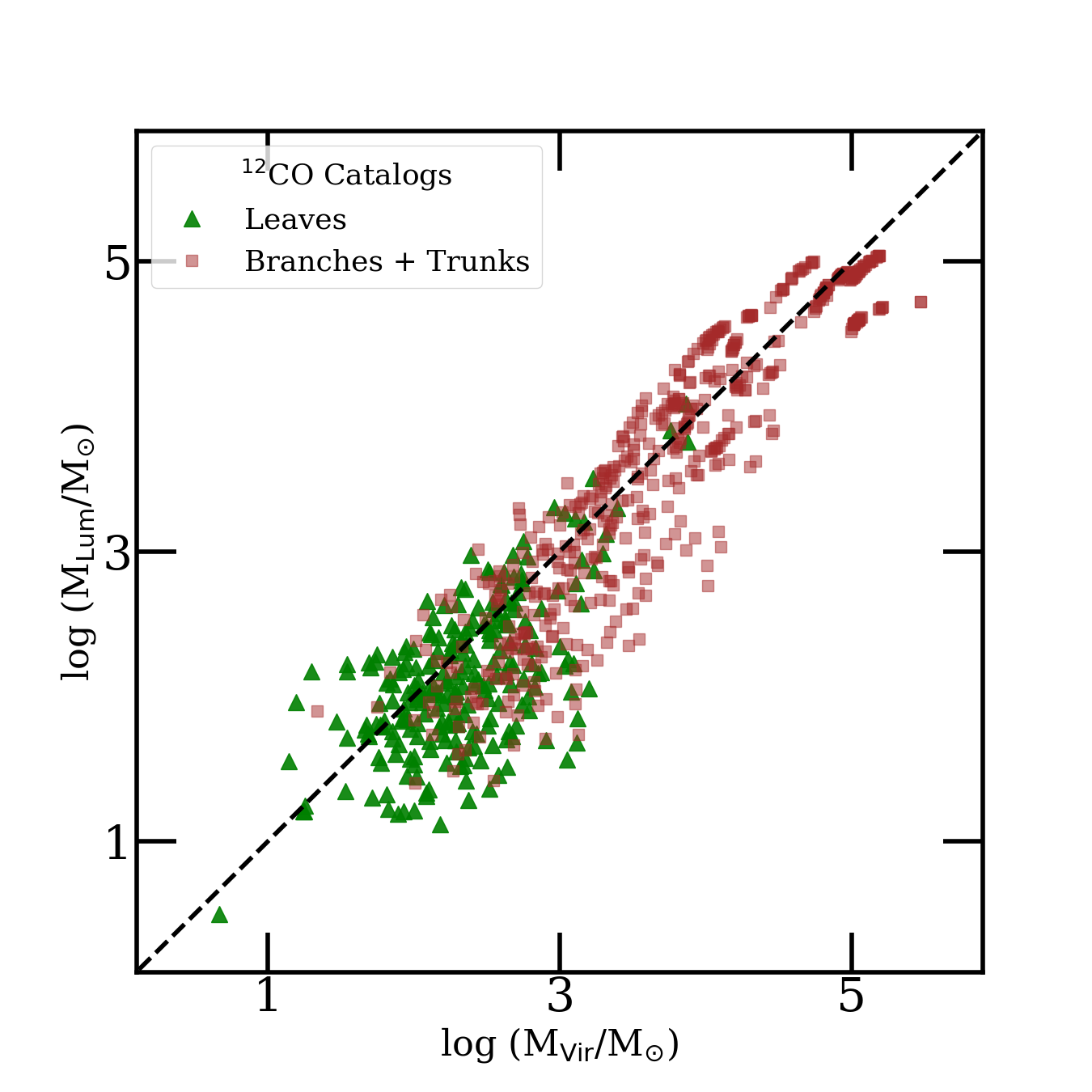} \\
	\includegraphics[width=9.45cm]{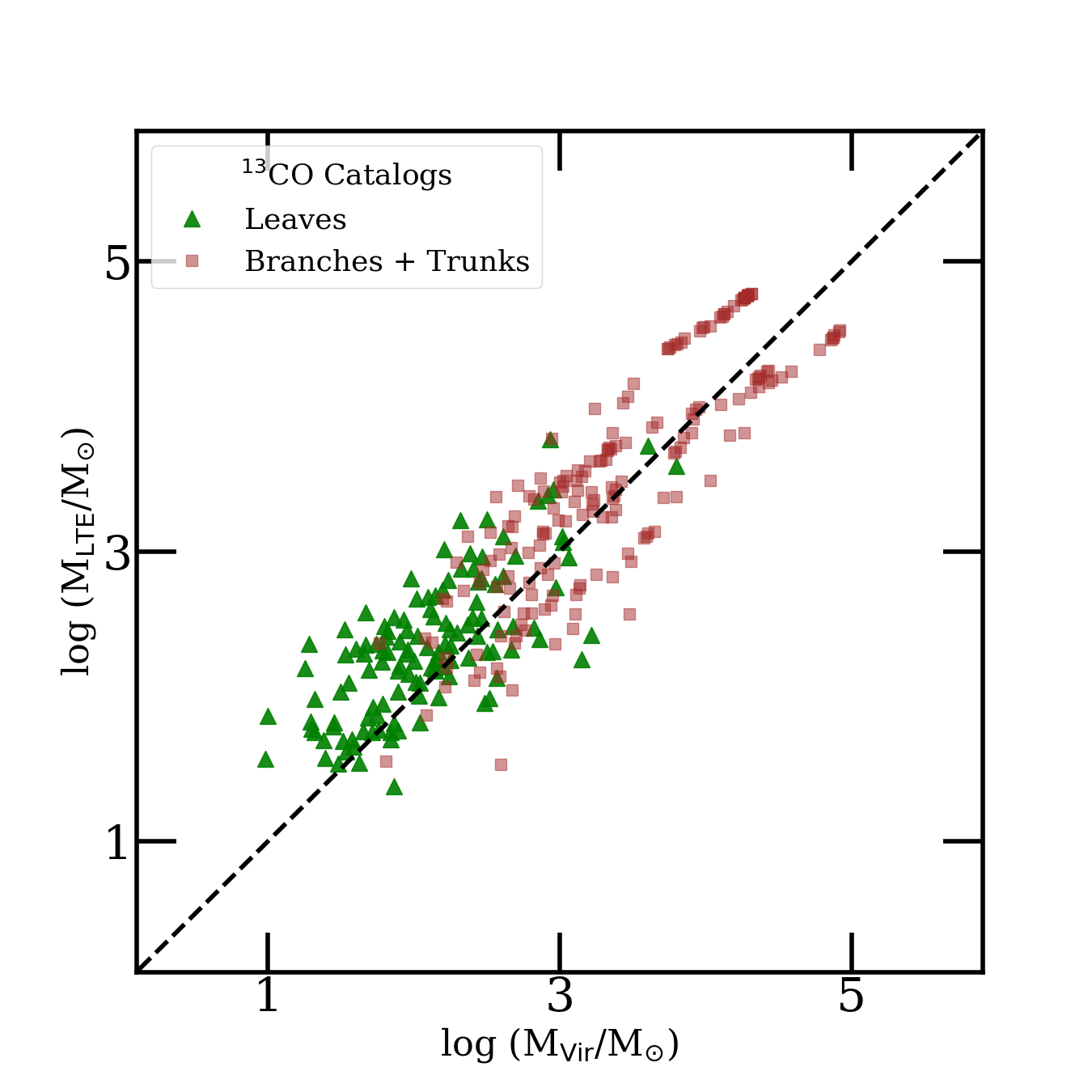} \hspace{-1.15cm}
	\includegraphics[width=9.45cm]{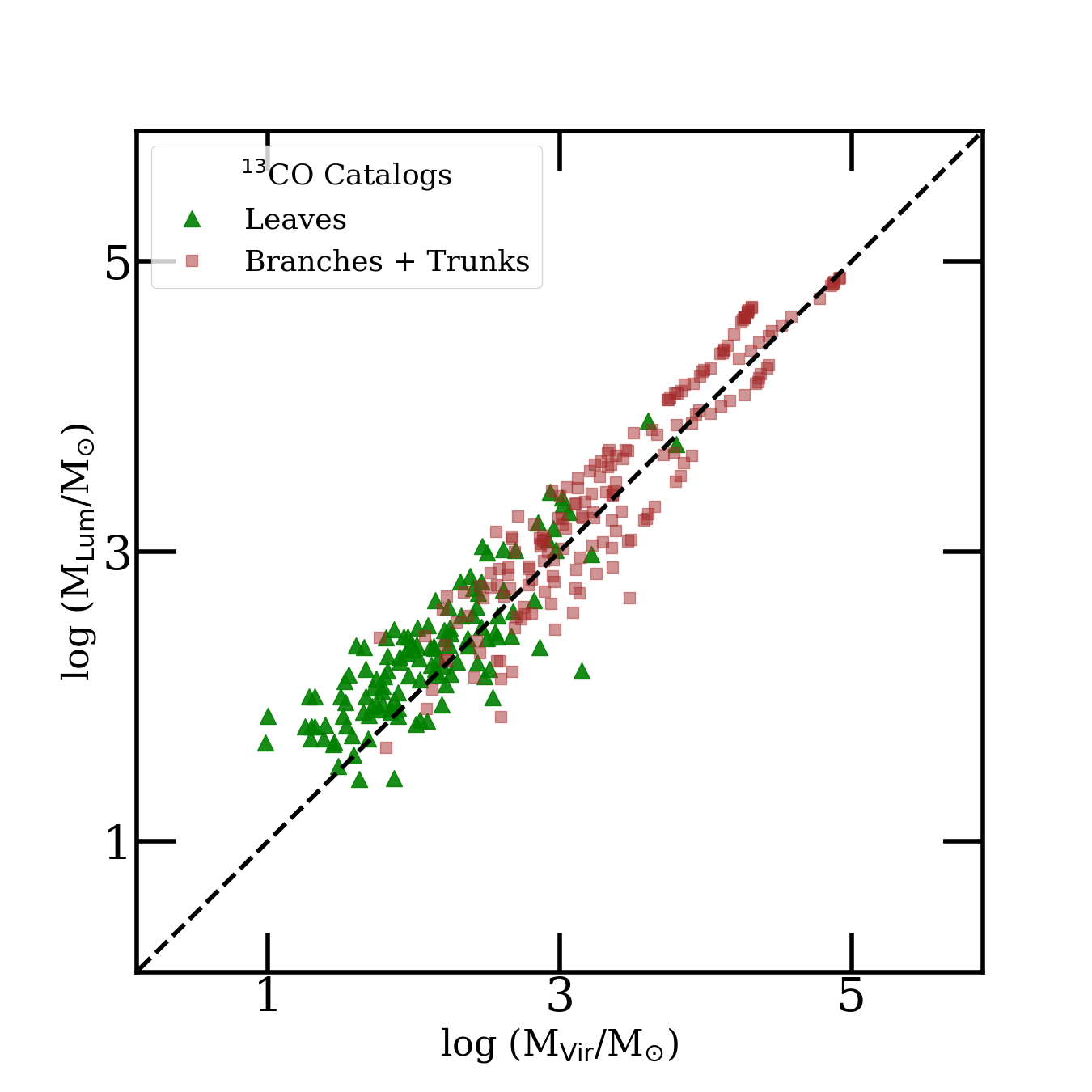}
	\caption{Comparison of the three mass estimates M$_{\rm Lum}$, M$_{\rm LTE}$, and M$_{\rm Vir}$ for clumps in the $^{12}$CO and $^{13}$CO catalogs as indicated in the panel legends. Leaves are represented by green triangles and branches and trunks by brown squares.}\label{fig-1213comp}
\end{figure}

In Figure~\ref{fig-1213comp}, we compare the three different mass estimates for the $^{12}$CO- and $^{13}$CO-detected clumps. The $^{12}$CO catalog (top panels) contains more clumps than the $^{13}$CO catalog (bottom panels), as expected. For both catalogs, there is less scatter between $M_{\rm Lum}$ and $M_{\rm Vir}$ (right panels) than between $M_{\rm LTE}$ and $M_{\rm Vir}$ (left panels), likely because both $^{12}$CO and $^{13}$CO measurements are required for $M_{\rm LTE}$ estimates, and the $^{13}$CO line is barely detected in many clumps. In any case, the correlations between the three mass estimates are close to linear in both catalogs (except for $M_{\rm LTE}$ vs. $M_{\rm Vir}$ at low masses in the $^{12}$CO catalog). In the following, we repeat our analysis with all three mass estimates in both the $^{12}$CO and $^{13}$CO catalogs as a guide to the uncertainties in our results.
\par
The masses of clumps in the \citet{Wong19} catalog range from below $10^2~M_{\odot}$ to above $10^5~M_{\odot}$, while their radii range from below 1~pc to above 10~pc. The typical surface density, volume density, and free-fall time of the clumps are $\Sigma=M/(\pi R^2)\sim 10^2~M_{\odot}~\mbox{pc}^{-2}$, $\rho=3M/(4\pi R^3) \sim 10^2~M_{\odot}~\mbox{pc}^{-3}$, and $t_{\rm ff}=(3\pi/32G\rho)^{1/2}\sim10^6$~yr, with only weak dependencies on mass (see Section~3.1 below). While these properties seem conducive to the onset of gravitational collapse and star formation, the clumps have not yet been systematically surveyed for direct evidence of recent star formation (young stellar objects, etc).
\section{Results} \label{sec:results}
\subsection{Mass-Radius Relation} \label{subsec:results-massradius}
\begin{figure}[ht]
	\centering
	\includegraphics[width=18.5cm]{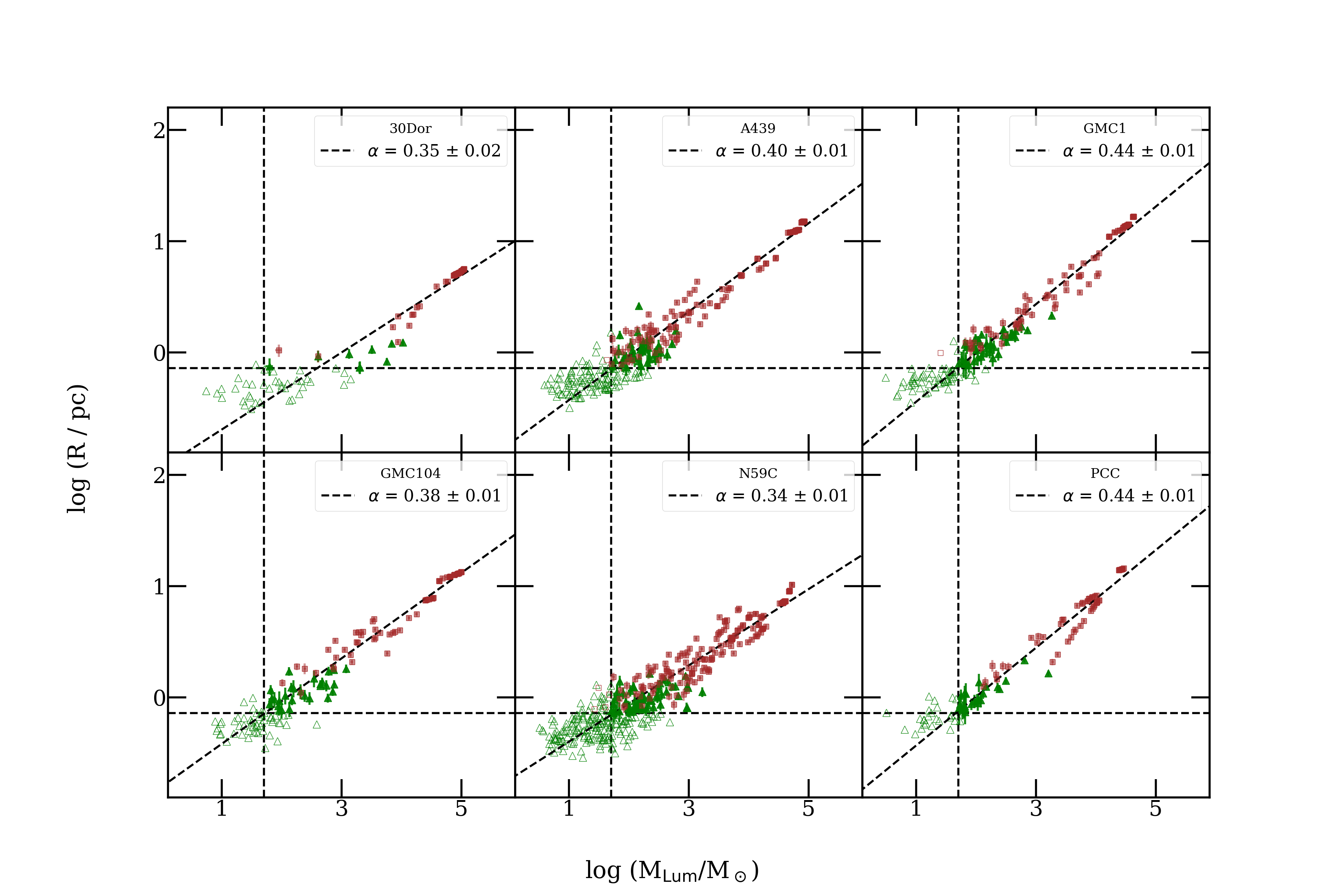}
	\caption{Mass-radius relations of clumps in the six GMCs based on the M$_{\rm Lum}$ mass estimate and $^{12}$CO catalog. Leaves are represented by green triangles and branches and trunks by brown squares. The best-fit indices $\alpha$ of the mass-radius relations are given in the panel legends. The horizontal and vertical dashed lines show the adopted completeness limits at $R = 0.72$~pc and log~$(M/M_\odot) = 1.75$.}\label{fig-massradius}
\end{figure}

In Figure~\ref{fig-massradius}, we present the mass-radius relation separately for clumps in the six GMCs from the $M_{\rm Lum}$ mass estimates in the $^{12}$CO catalog. Evidently, these relations all have similar power-law form, $R \propto M^{\alpha}$. We determine the best-fit values of $\alpha$ and their standard errors from the linear regression $\log R=\alpha \log M + {\rm constant}$ for clumps with $R\geq 0.72$~pc (the completeness limit adopted by \citealt{Wong19}) and log$(M/M_\odot) \geq 1.75$ (the completeness limit we adopt for the mass functions in Section~\ref{subsec:results-massfunc}). The results of these individual fits for the six GMCs are displayed at the upper right of the panels in Figure~2. We obtain nearly identical results for the other mass estimates ($M_{\rm LTE}$ and $M_{\rm Vir}$) and for the $^{13}$CO catalog. 
\par
The variations in $\alpha$ from one GMC to another ($\sim0.04$) appear to be several times larger than the small statistical uncertainty in each value of $\alpha$ (0.01-0.02). This may indicate that there are real variations in the mass-radius relation of clumps among different GMCs. \citet{Wong19} speculate that analogous variations in the size-linewidth relation may be caused by different levels of stellar feedback in the GMCs. Testing this conjecture will require a full census of the energy and momentum output of all the massive young stars in each GMC, a worthwhile project for the future. Alternatively, the apparent variations in $\alpha$ may simply reflect small systematic (non-statistical) errors, such as those inherent to low-density molecular tracers and to any method for identifying clumps. 
\par
In the following, we present results only for the combined sample of clumps in all six GMCs. The best-fit values of $\alpha$ and their standard errors for the combined clump sample are listed in Table~\ref{tab:alpha} for each of the three mass estimates and for the $^{12}$CO and $^{13}$CO catalogs. These six indices are all remarkably similar, differing from each other only by the small statistical uncertainties ($\pm0.01$). The median and mean indices are $\alpha=0.36$. To allow for systematic errors, at least roughly, as discussed above, we adopt $\alpha=0.36\pm0.03$. This value of $\alpha$ indicates that the volume densities of clumps are nearly independent of their masses ($\alpha=1/3$), a slightly weaker dependence of $R$ on $M$ than for constant surface density ($\alpha=1/2$).
\begin{deluxetable*}{ccc}[ht]
 \tablecaption{Index $\alpha$ of the Mass-Radius Relation, $R \propto M^{\alpha}$\label{tab:alpha}}
 \tablehead{\colhead{Mass} & \colhead{$^{12}$CO} & \colhead{$^{13}$CO}}
	\startdata
	$M_{\rm Lum}$ & $0.35\pm0.01$ & $0.36\pm0.01$ \\
 $M_{\rm Vir}$ & $0.36\pm0.01$ & $0.35\pm0.01$ \\
 $M_{\rm LTE}$ & $0.35\pm0.01$ & $0.37\pm0.01$ \\
	\enddata
	\tablenotetext{}{Regression fits for leaves, branches, and trunks with $R \geq 0.72$ pc and log $(M/M_\odot) \geq 1.75$.}
\end{deluxetable*}
\par
It is interesting to compare the mass-radius relation derived here for LMC clumps identified by dendogram analysis of CO maps with that derived for Milky Way clumps identified by various methods. FKM found $\alpha=0.38\pm0.02$ for a composite sample of clumps selected for their star-formation activity and with measurements of CS, C$^{17}$O, or 1.2 mm dust emission from three independent surveys \citep{Shirley03, Faundez04, Fontani05}. Our value of $\alpha$ for LMC clumps is nearly identical to the FKM value for Milky Way clumps. Our result also falls within the range $0.3 \la \alpha \la 0.6$ found in other studies of clumps in the Milky Way \citep{Wong08, Wu10, Roman-Duval10, Urquhart18}.
\par
Many of the leaves from the dendogram analysis, with $R \sim 1$~pc, are only marginally resolved with the available ALMA observations.  Nevertheless, it appears from Figure~\ref{fig-massradius} that they follow the same mass-radius relation as the branches and trunks, consistent with a scale-free hierarchy of turbulence within the GMCs \citep{Wong19}.  Thus, even for the leaves alone, we rely on the index $\alpha=0.36 \pm 0.03$ derived for the full range of clump masses and sizes (above the completeness limits).  As noted above, this is consistent with the mass-radius relation of clumps in the Milky Way, which are resolved down to $R \sim 0.1$~pc (see Figure~1 of FKM).
\subsection{Mass Function} \label{subsec:results-massfunc}
\par
We derive the mass function of LMC clumps from the leaves of the \citet{Wong19} dendogram analysis. There are two justifications for this. (1) Because the leaves are defined by local peaks in the intensity maps, they are the most likely sites of cluster formation. (2) Because the branches and trunks contain the leaves in a nested hierarchy, including them would lead to inconsistencies in accounting for the numbers and masses of clumps and hence their mass function. We stress that, in the absence of a complete theory of clump and cluster formation, there is no unique or even generally accepted choice of exactly which ISM structures to identify as protoclusters. As we show here, the choice of dendogram leaves leads to a consistent picture of cluster formation.
\par
\begin{figure}[ht]
	\centering
	\includegraphics[width=12.5cm]{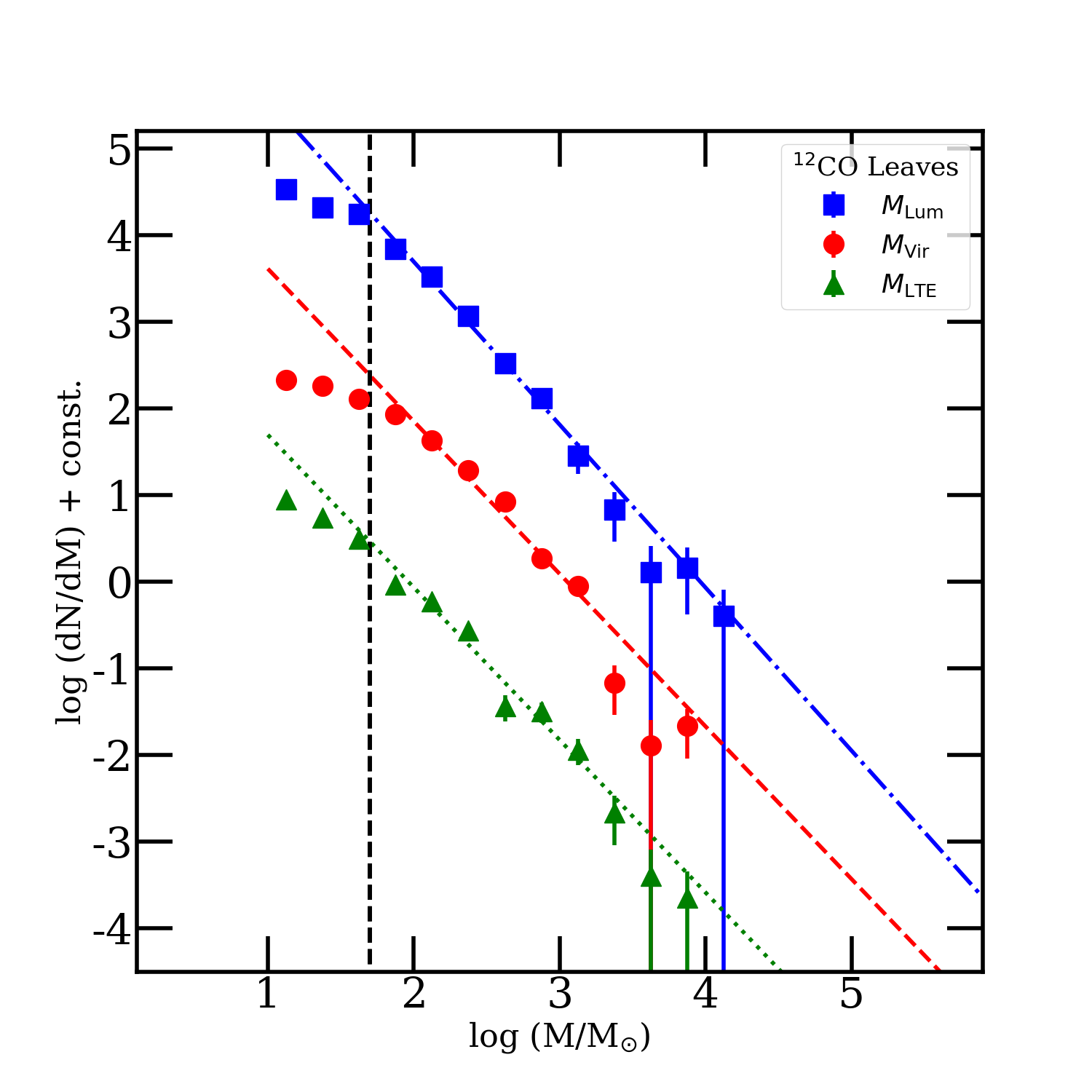}
	\caption{Mass functions of leaves in the $^{12}$CO catalog for the M$_{\rm Lum}$, M$_{\rm LTE}$, and M$_{\rm Vir}$ mass estimates in equal logarithmic bins. The vertical normalizations have been shifted for clarity. The diagonal lines are maximum-likelihood fits of power laws to the unbinned masses with best-fit indices $\beta$ listed in Table~\ref{tab:beta}. The vertical dashed line shows the adopted completeness limit at $\log (M/M_\odot) = 1.75$ (see text).}\label{fig-massfunction}
\end{figure}

 
In Figure~\ref{fig-massfunction}, we present the mass functions of leaves in the combined GMC sample with the three mass estimates in the $^{12}$CO catalog. These have similar power-law form, $dN/dM \propto M^{\beta}$, above our adopted completeness limit at log $(M/M_\odot)=1.75$. Below this limit, the mass functions begin to fall significantly below the extrapolated power laws (by a factor of two in the case of $M_{\rm Vir}$). We derive the best-fit values of $\beta$ and their standard errors from maximum-likelihood fits to the unbinned masses above the completeness limit (see \cite{Mok19} for details of the method). These are listed in Table~\ref{tab:beta} for the three mass estimates and both the $^{12}$CO and $^{13}$CO catalogs. The fitted values of $\beta$ are not sensitive to the adopted lower mass limit so long as it lies near or above log~$(M/M_\odot)=1.75$. The median and mean indices in Table~\ref{tab:beta} are $\beta=-1.79$ and $\beta=-1.81$, respectively. Thus, we adopt $\beta=-1.8\pm0.1$ to allow for possible systematic errors of the kinds we discussed previously for $\alpha$ (Section~\ref{subsec:results-massradius}).
\begin{deluxetable*}{ccc}[ht]
 \tablecaption{Index $\beta$ of the Mass Function, $dN/dM \propto M^{\beta}$\label{tab:beta}}
 \tablehead{\colhead{Mass} & \colhead{$^{12}$CO} & \colhead{$^{13}$CO}}
	\startdata
	$M_{\rm Lum}$ & $-1.88_{-0.03}^{+0.02}$ & $-1.81_{-0.06}^{+0.04}$ \\
 $M_{\rm Vir}$ & $-1.76_{-0.01}^{+0.04}$ & $-1.87_{-0.07}^{+0.08}$ \\
 $M_{\rm LTE}$ & $-1.76_{-0.06}^{+0.04}$ & $-1.77_{-0.07}^{+0.08}$ \\
	\enddata
	\tablenotetext{}{Maximum-likelihood fits for leaves with log $(M/M_\odot) \geq 1.75$.}
\end{deluxetable*}
\par
\begin{figure}[ht]
	\centering
	\includegraphics[width=12.5cm]{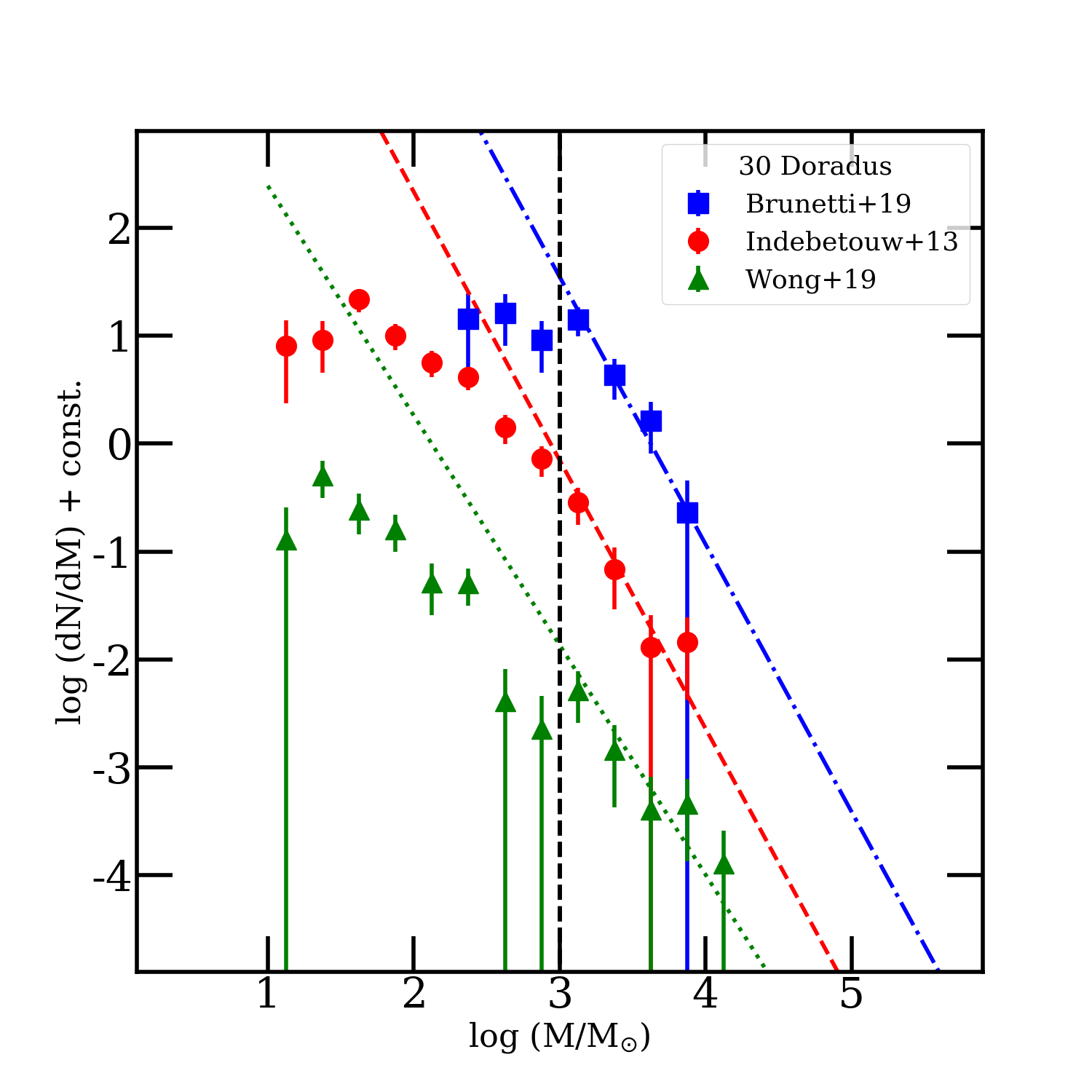}
	\caption{Mass functions of clumps in the 30-Dor GMC from three independent catalogs: gas/dendrogram \citep{Wong19}, gas/clumpfind \citep{Indebetouw13}, and dust/clumpfind \citep{Brunetti19}. The vertical normalizations have been shifted for clarity.
	The diagonal lines are maximum-likelihood fits of power laws to the unbinned masses with best-fit indices $\beta = -2.13_{-0.38}^{+0.42}$ (gas/dendogram), $\beta = -2.49_{-0.39}^{+0.51}$ (gas/clumpfind), and $\beta = -2.48_{-0.33}^{+0.37}$ (dust/clumpfind). The vertical dashed line shows the adopted completeness limit of the shallowest survey at $\log (M/M_\odot) = 3.0$.}\label{fig-massfunctioncomp}
\end{figure}

There are two previous determinations of the mass function of clumps in 30 Dor, one of the GMCs included in the present study, from the same ALMA data analyzed by \cite{Wong19}. The first is based on clumps identified by clumpfind in the $^{12}$CO map \citep{Indebetouw13} and the second on clumps identified by clumpfind in the 1.3~mm and 3.2~mm dust continuum maps \citep{Brunetti19}. We plot these mass functions in Figure~\ref{fig-massfunctioncomp}, along with the one derived from clumps identified as dendogram leaves in 30 Dor in the \citet{Wong19} $^{12}$CO catalog. Evidently, the high-mass end of these functions (log $(M/M_\odot) \ge 3$) have similar power-law form with $\beta \approx-2$ or slightly steeper, but the number of clumps is small and the uncertainties large. Thus, all we can conclude from this comparison is that the mass functions derived from gas and dust maps and dendogram and clumpfind analyses are consistent with each other.
\par
We can make a more definitive comparison between the mass function derived here for LMC clumps identified as dendogram leaves of CO maps and that for Milky Way clumps identified by various methods. There are numerous studies of clumps in the Milky Way \citep[e.g][]{Munoz07, Wong08, Schlingman11, Pekruhl13, Urquhart14, Moore15}. The mass functions of clumps derived in these studies all have indices in the range $-1.9 \leq \beta \leq -1.6$, with a mean and standard deviation $\beta=-1.8\pm0.1$, identical to our result for LMC clumps.
\section{Interpretation} \label{sec:interpretation}
\par
We now check for consistency between the FKM feedback model, the results of the previous section, and the observed mass function of young clusters in the LMC. The FKM model is based on the assumption that stars will continue to form in a protocluster until they have injected enough energy or momentum to remove the remaining gas, thus ending further star formation and fixing the SFE. This condition, applied to protoclusters of different masses, then determines the mass function of the resulting clusters (with index $\beta_{\rm cluster}$) in terms of the mass function and mass--radius relation of their antecedent clumps (with indices $\beta_{\rm clump}$ and $\alpha_{\rm clump}$, respectively). In the analytical FKM model, the SFE depends on the masses of the protoclusters mainly through their surface densities, a result confirmed by recent hydrodynamical simulations \citep{Grudic18, Kim18}. 
\par
Feedback in protoclusters potentially involves non-linear combinations of protostellar outflows, main-sequence winds, photoionized gas, radiation pressure on dust, and supernovae. However, the gas removal in this complex realistic situation is likely bracketed by that in the simple idealized cases in which the feedback is energy driven or momentum driven, corresponding to minimum and maximum radiative losses, respectively. The predictions of the FKM model in these limiting cases are:
\begin{subequations}\label{eqn-1}
 \begin{align*}
 \beta_{\rm cluster} &= \frac{2(\beta_{\rm clump} + \alpha_{\rm clump} - 1)}{5(1 - \alpha_{\rm clump})} \text{\hspace{2cm}(energy driven)~~~~~~~~~~~~~~~~~(1)} \\ \\
 \beta_{\rm cluster} &= \frac{2\beta_{\rm clump} + \alpha_{\rm clump} - 1}{4(1 - \alpha_{\rm clump})} \text{\hspace{2cm}(momentum driven)~~~~~~~~~~~~~(2)}.
 \end{align*}
\end{subequations}
\par
We compute the index of the cluster mass function from these equations with the clump parameters derived in the previous section: $\alpha_{\rm clump} = 0.36 \pm 0.03$ and $\beta_{\rm clump} = -1.8 \pm 0.1$. The results are $\beta_{\rm cluster} = -1.53 \pm 0.08$ (energy driven) and $\beta_{\rm cluster} = -1.66 \pm 0.10$ (momentum driven). Evidently, the predicted $\beta_{\rm cluster}$ is not sensitive to whether the feedback in protoclusters is closer to the energy-driven or momentum-driven limits.
\par 
There have been several determinations of the mass function of young clusters in the LMC, leading to $\beta_{\rm cluster} = -1.8 \pm 0.2$ \citep[][and references therein]{Chandar10a}. Most recently, we derived $\beta_{\rm cluster} = -1.67 \pm 0.05$ by maximum-likelihood fit of a pure power law to the unbinned masses of LMC clusters with $M \ge 300~M_\odot$ and $\tau \le 10~{\rm Myr}$ \citep{Mok20}, i.e., the same method we use here to derive $\beta_{\rm clump}$. This result is based on the \citet{Hunter03} catalog, which covers most of the LMC, with revised mass and age estimates from \citet{Chandar10a}. For momentum-driven feedback, the value of $\beta_{\rm cluster}$ predicted by the FKM model agrees perfectly with the observed value. The model predicts a slightly shallower mass function for energy-driven feedback, but the agreement with observation is still satisfactory given the uncertainties (with $\Delta\beta_{\rm cluster} = 0.14\pm 0.09$).
\par 
In a previous paper, we made similar comparisons between the predicted and observed $\beta_{\rm cluster}$ for the cluster and GMC populations in six nearby galaxies \citep{Mok20}. For five of these galaxies, we found $\beta_{\rm cluster} \approx \beta_{\rm GMC} \approx -2$ (within $\pm 0.3$) and good agreement with the FKM model. The Milky Way has $\beta_{\rm cluster} \approx \beta_{\rm clump} \approx \beta_{\rm GMC} \approx -2$ (within $\pm 0.3$), consistent with the FKM model for both clumps and GMCs (\citealt{Lada03}; FKM; \citealt{Rice16}). For the LMC, however, the GMC mass function, with $\beta_{\rm GMC} \approx -2.6$ \citep{Wong11}, appears to be significantly steeper than both the cluster and clump mass functions, with $\beta_{\rm cluster} \approx \beta_{\rm clump} \approx -1.7$ \citep[][this work]{Chandar10a}. Thus, while the FKM model is successful for clumps in the LMC, it is not for GMCs, unlike the situation in the Milky Way, the only other galaxy for which this clump-GMC comparison is possible. This is an intriguing puzzle, worthy of further study.
\par
The FKM feedback model provides a physical explanation for the observed index $\beta_{\rm cluster}$ of the cluster mass function in terms of the observed indices $\beta_{\rm clump}$ and $\alpha_{\rm clump}$ of the clump mass function and mass-radius relation. In this model, the properties of the clumps are regarded as ``initial conditions" for the clusters. In a more complete theory, the formation and properties of the clumps would also need to be explained. A variety of idealized analytical models and hydrodynamical simulations suggest that the observed indices $\beta_{\rm clump} \approx -2.0$ and $\alpha_{\rm clump} \approx 0.4$ could plausibly arise from turbulence and/or related processes (fragmentation, coalescence, accretion) in a hierarchically structured ISM (\citealt{Elmegreen96, Fleck96, Guszejnov18, Kuznetsova18}, but see \citealt{Guszejnov20} for counterexamples). This is a topic that will continue to benefit from hydrodynamical simulations of increasing numerical resolution and astrophysical realism. 
\section{Conclusions} \label{sec:conclusion}
\par
In summary, we have derived the mass-radius relation and mass function of clumps in the dendrogram-based catalog compiled by \citet{Wong19} from ALMA $^{12}$CO and $^{13}$CO maps of six GMCs in the LMC. The Magellanic Clouds are the only external galaxies for which this type of analysis is possible at the necessary spatial resolution ($\sim1$~pc). Our main goals in this study were to compare the mass-radius relation and mass function of clumps in a large, uniform, and representative (but incomplete) sample in the LMC with those in the Milky Way and to test the FKM feedback model for consistency with the observed properties of clumps and clusters in the LMC. Our main conclusions are the following. 
\begin{enumerate}
 \item The mass-radius relation of LMC clumps has power-law form, $R \propto M^{\alpha}$, with index $\alpha = 0.36\pm0.03$ over the mass range $10^2 M_\odot \la M \la 10^5 M_\odot$. This is very similar to the mass-radius relation of clumps in the Milky Way.
 \item The mass function of LMC clumps has power-law form, $dN/dM \propto M^{\beta}$, with index $\beta = -1.8 \pm 0.1$ over the mass range $10^2 M_\odot \la M \la 10^4 M_\odot$. This is very similar to the mass function of clumps in the Milky Way.
 \item With these values of $\alpha$ and $\beta$ for the clumps, the predicted index for the mass function of young LMC clusters from the FKM model is $\beta \approx -1.7$, in excellent agreement with the observed index. This again is very similar to the situation in the Milky Way. 
\end{enumerate}
\par
To put the work presented here into a broader context, we note that the lives of clusters can be loosely divided into three phases, based on their locations in a galaxy and the dominant physical processes responsible for their dissolution. In the first phase, as stellar feedback expels the gas from protoclusters, their gravitational binding is weakened, thus allowing some or even all of their stars to escape (protocluster phase). In the second phase, clusters that remain in the galactic disk are dissolved primarily by stellar escape driven by tidal encounters with passing GMCs (disk-cluster phase). In the third phase, clusters whose orbits have been scrambled into the galactic halo by encounters and mergers with other galaxies are dissolved primarily by stellar escape driven by internal two-body relaxation (halo-cluster phase). The cluster population is decimated during each of these phases, leaving only a few surviving globular clusters after a Hubble time. During the protocluster and disk-cluster phases, the amount and rate of mass loss are essentially independent of mass, thus preserving the power-law shape and index $\beta \approx -2$ of the mass function while its normalization declines, as observed (FKM; \citealt{Fall12}). During the halo-cluster phase, low-mass clusters dissolve faster than high-mass clusters, and the mass function develops a bend at $M_p \sim 10^5 M_{\odot}$ after $\sim 10^{10}$~yr, also as observed \citep{Fall01, McLaughlin08}. 
\section*{Acknowledgements} \label{sec:acknowledgements}
\par
We thank Nathan Brunetti, Mark Krumholz, Chris Matzner, and Tony Wong for helpful comments on the submitted version of this paper. R.C. acknowledges support from NSF grant 1517819.
\bibliography{master}
\end{document}